\documentclass[prl,twocolumn,nofootinbib]{revtex4}
\usepackage{bm,amsmath,amssymb,graphicx}
\usepackage{pgfplots}
\pgfplotsset{width=10cm,compat=1.9}

\begin{document}

\title{Analysis of big bounce in Einstein--Cartan cosmology}

\author{Jordan L. Cubero}
\altaffiliation{jcube1@unh.newhaven.edu}

\author{Nikodem J. Pop{\l}awski}
\altaffiliation{NPoplawski@newhaven.edu}

\affiliation{Department of Mathematics and Physics, University of New Haven, 300 Boston Post Road, West Haven, CT 06516, USA}

\begin{abstract}
We analyze the dynamics of a homogeneous and isotropic universe in the Einstein--Cartan theory of gravity.
The spin of fermions produces spacetime torsion that prevents gravitational singularities and replaces the big bang with a nonsingular big bounce.
We show that a closed universe exists only when a particular function of its scale factor and temperature is higher than some threshold value, whereas an open and a flat universes do not have such a restriction.
We also show that a bounce of the scale factor is double: as the temperature increases and then decreases, the scale factor decreases, increases, decreases, and then increases. 
\end{abstract}
\maketitle

{\it Introduction.}
The Einstein--Cartan (EC) theory of gravity provides the simplest mechanism generating a nonsingular big bounce, without unknown parameters or hypothetical fields \cite{avert}.
EC is the simplest and most natural theory of gravity with torsion, in which the Lagrangian density for the gravitational field is proportional to the Ricci scalar, as in general relativity \cite{KS, EC}.
The consistency of the conservation law for the total (orbital plus spin) angular momentum of fermions in curved spacetime with the Dirac equation allowing the spin-orbit interaction requires that the antisymmetric part of the affine connection, the torsion tensor \cite{Schr}, is not constrained to zero \cite{req}.
Instead, torsion is determined by the field equations obtained from the stationarity of the action under variations of the torsion tensor.
In EC, the spin of fermions is the source of torsion.

The multipole expansion \cite{Pap} of the conservation law for the spin tensor in EC gives a spin tensor which describes fermionic matter as a spin fluid (ideal fluid with spin) \cite{NSH}.
Once the torsion is integrated out, EC reduces to general relativity with an effective spin fluid as a matter source \cite{EC}.
The effective energy density and pressure of a spin fluid are given by $\tilde{\epsilon}=\epsilon-\alpha n_\textrm{f}^2$ and $\tilde{p}=p-\alpha n_\textrm{f}^2$, where $\epsilon$ and $p$ are the thermodynamic energy density and pressure, $n_\textrm{f}$ is the number density of fermions, and $\alpha=\kappa(\hbar c)^2/32$ with $\kappa=8\pi G/c^4$ \cite{HHK,spin}.
The negative corrections from the spin-torsion coupling generate gravitational repulsion which prevents the formation of gravitational singularities and replaces the big bang with a nonsingular bounce, at which the universe transitions from contraction to expansion.
These corrections lead to a violation of the strong energy condition by the spin fluid when $\epsilon+3p-4\alpha n_\textrm{f}^2$ drops below 0, thus evading the singularity theorems \cite{HHK}.
Accordingly, this violation could be thought of as the cause of the bounce.

The dynamics of the universe filled with a spin fluid in EC has been studied in \cite{spin}.
The expansion of the closed universe with torsion and quantum particle production shortly after a bounce is almost exponential for a finite period of time, explaining inflation \cite{ApJ}.
Depending on the particle production rate, the universe may undergo several bounces until it produces enough matter to reach a size where the cosmological constant starts cosmic acceleration.
This expansion also predicts the cosmic microwave background radiation parameters that are consistent with the Planck 2015 observations \cite{Planck2015}, as was shown in \cite{SD}.

The spin-fluid description of matter is the particle approximation of the multipole expansion of the integrated conservation laws in EC.
The particle approximation for Dirac fields, however, is not self-consistent \cite{nons}.
The spin-fluid description also violates the cosmological principle \cite{viol}.
On the other hand, the Dirac form of the spin tensor for fermionic matter, which follows directly from the Dirac Lagrangian using the Sciama-Kibble variation with respect to the torsion tensor \cite{KS,spinor}, is consistent with the cosmological principle \cite{consist}.
The minimal coupling between the torsion tensor and Dirac fermions also averts the big-bang singularity \cite{Bounce,also}.
Other forms of the torsion tensor in cosmology are investigated in \cite{other}.

The spin-torsion coupling modifies the Dirac equation, adding a term that is cubic in spinor fields \cite{HD}.
As a result, fermions must be spatially extended \cite{nons}, which could eliminate infinities arising in Feynman diagrams involving fermion loops.
In the presence of torsion, the four-momentum operator components do not commute and thus the integration in the momentum space in Feynman diagrams must be replaced with the summation over the discrete momentum eigenvalues.
The resulting sums are finite: torsion naturally regularizes ultraviolet-divergent integrals in quantum electrodynamics \cite{toreg}.
Torsion may also explain the matter-antimatter asymmetry \cite{anti} and the cosmological constant \cite{cosmo}.

The analysis in \cite{Bounce} considered a closed, homogeneous, and isotropic universe in EC.
However, the calculations of the maximum temperature and the minimum scale factor at a bounce neglected the factor $k=1$ in the Friedmann equations (which is justified during and after inflation but not at a bounce before inflation), de facto considering a flat universe.
In addition, the time-dependence of the scale factor appeared to have a cusp-like behavior at the bounce.
In this article, we refine those calculations by taking $k$ into account and analyzing the turning points of the universe for all three cases: $k=1$ (closed universe), $k=0$ (flat universe), and $k=-1$ (open universe).
We discover that a closed universe exists only when some function of the scale factor and temperature is higher than a particular threshold, whereas an open and flat universes are not restricted by this condition.
Accordingly, a closed universe forms in a region of space within a trapped null surface when this threshold is reached.
Such a region could be the interior of a black hole \cite{BH}.
Torsion therefore may explain the origin of our universe \cite{ApJ}.

{\it Dynamics of scale factor and temperature.}
If we assume that the universe is homogeneous and isotropic, then it is described by the Friedmann-Lema\^{i}tre-Robertson-Walker metric, which is given in the isotropic spherical coordinates by
\begin{equation}
ds^2=c^2 dt^2-\frac{a^2(t)}{(1+kr^2/4)^2}(dr^2+r^2 d\theta^2+r^2\sin^2\theta\,d\phi^2),
\label{metric}
\end{equation}
where $a(t)$ is the scalar factor as a function of the cosmic time $t$ \cite{LL2}.
The Einstein field equations for this metric become the Friedmann equations:
\begin{equation}
\frac{\dot{a}^2}{c^2}+k=\frac{1}{3}\kappa\tilde{\epsilon} a^2
\label{energy}
\end{equation}
and
\begin{equation}
\frac{\dot{a}^2+2a\ddot{a}}{c^2}+k=-\kappa\tilde{p}a^2,
\label{Fri}
\end{equation}
where a dot denotes the derivative with respect to $t$.
The effective energy density and pressure for a Dirac field are given by \cite{spinor,Bounce}.
\begin{equation}
\tilde{\epsilon}=\epsilon-\alpha n_\textrm{f}^2,\quad\tilde{p}=p+\alpha n_\textrm{f}^2,
\end{equation}
where
\begin{equation}
\alpha=\frac{9}{16}\kappa(\hbar c)^2.
\end{equation}
Multiplying the first Friedmann equation by $a$ and differentiating over time, and subtracting from it the second Friedmann equation multiplied by $\dot{a}$ gives an equation that has the form of the first law of thermodynamics for an adiabatic universe:
\begin{equation}
\frac{d}{dt}\bigl((\epsilon-\alpha n_\textrm{f}^2)a^3\bigr)+(p+\alpha n_\textrm{f}^2)\frac{d}{dt}(a^3)=0,
\end{equation}
which gives
\begin{equation}
a^3 d\epsilon - 2\alpha a^3 n_\textrm{f}\,dn_\textrm{f} + (\epsilon + p)d(a^3) = 0.
\label{cons}
\end{equation}

The matter in the early universe is ultrarelativistic.
If we assume kinetic equilibrium, then we can use $\epsilon=h_\star T^4$, $p=\epsilon/3$, and $n_\textrm{f}=h_{n\textrm{f}}T^3$, where $T$ is the temperature of the universe, $h_\star=(\pi^2/30)(g_\textrm{b}+(7/8)g_\textrm{f})k_\textrm{B}^4/(\hbar c)^3$, and $h_{n\textrm{f}}=(\zeta(3)/\pi^2)(3/4)g_\textrm{f}k_\textrm{B}^3/(\hbar c)^3$ \cite{Ric}.
The quantities $g_\textrm{b}$ and $g_\textrm{f}$ are the numbers of spin states for all elementary bosons and fermions, respectively.
For standard-model particles, $g_\textrm{b}=29$ and $g_\textrm{f}=90$.
In the presence of spin and torsion, the first Friedmann equation is therefore \cite{Bounce}
\begin{equation}
\frac{{\dot{a}}^2}{c^2}+k=\frac{1}{3}\kappa(h_\star T^4-\alpha h_{n\textrm{f}}^2 T^6)a^2.
\label{Nikoeq1}
\end{equation}
The first law of thermodynamics (\ref{cons}) gives
\begin{equation}
\frac{dT}{T}-\frac{3\alpha h_{n\textrm{f}}^2}{2h_\star}T\,dT+\frac{da}{a}=0
\end{equation}
or
\begin{equation}
\frac{dT}{T} -\frac{TdT}{T^2_\textrm{cr}}+\frac{da}{a} = 0,
\label{aT1}
\end{equation}
where
\begin{equation}
T_\textrm{cr}=\Bigl(\frac{2h_\star}{3\alpha h_{n\textrm{f}}^2}\Bigr)^{1/2}=2.218\times10^{31}\,\textrm{K}
\end{equation}
is the critical temperature.
Equation (\ref{aT1}) determines the scale factor $a$ as a function of temperature $T$ and can be integrated to
\begin{equation}
a(T)=\frac{C^\ast}{T}e^\frac{T^2}{2T^2_\textrm{cr}},
\label{aT2}
\end{equation}
where $C^\ast$ is an integration constant.
This constant is positive because the scale factor and temperature are positive.

We define nondimensional quantities:
\begin{eqnarray}
& & x=\frac{T}{T_\textrm{cr}}, \label{nont} \\
& & y=\frac{a}{a_\textrm{cr}}, \\
& & \tau=\frac{ct}{a_\textrm{cr}},
\end{eqnarray}
where
\begin{equation}
a_\textrm{cr}=\frac{27}{8}\hbar c\Bigl(\frac{\alpha h_{n\textrm{f}}^4}{h_\star^3}\Bigr)^{1/2}=6.661\times10^{-35}\,\textrm{m}.
\end{equation}
Henceforth, we will use a dot to denote the derivative with respect to the new time coordinate $\tau$. 
Equation (\ref{Nikoeq1}) can be written, using these quantities, as
\begin{equation}
\dot{y}^2+k=(3x^4-2x^6)y^2.
\label{Jordan1}
\end{equation}
Equation (\ref{aT2}) can be written as
\begin{equation}
y(x)=\frac{C}{x}e^{x^2/2},
\label{xy}
\end{equation}
where $C$ is a positive constant, related to $C^\ast$ by
\begin{equation}
C=\frac{C^\ast}{a_\textrm{cr}T_\textrm{cr}}.
\end{equation}
Substitution of $y(x)$ in Eq. (\ref{xy}) into Eq. (\ref{Jordan1}) gives
\begin{equation}
\dot{y}^2+k=C^2(3x^2 - 2x^4)e^{x^2}.
\label{Jordan2}
\end{equation}

\begin{figure}
\centering
\includegraphics[width=0.4\textwidth]{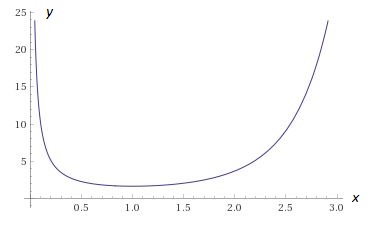}
\caption{The nondimensionalized scale factor $y$ as a function of the nondimensionalized temperature $x$ for $C=1$.
The minimum value of $y$ is positive: the universe is nonsingular.
}
\label{Fig1}
\end{figure}

The function $y(x)$, given by Eq. (\ref{xy}), has a minimum at $x=1$ ($T=T_\textrm{cr}$), and its value is
\begin{equation}
y_\textrm{min}=Ce^{1/2}.
\label{mini}
\end{equation}
This minimum determines the minimum value of the scale factor:
\begin{equation}
a_\textrm{min}=Ce^{1/2}a_\textrm{cr}>0.
\end{equation}
Consequently, the scale factor $a$ cannot reach zero.
The universe with spin and torsion is nonsingular for all three cases of $k$ and for all values of $C$.
Figure \ref{Fig1} shows the function (\ref{xy}) for $C=1$.

\begin{figure}
\centering
\includegraphics[width=0.4\textwidth]{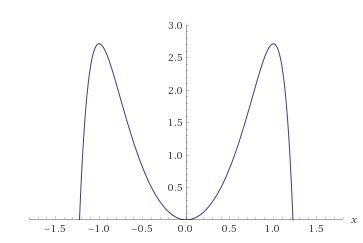}
\caption{A curve representing the function on the left-hand side of Eq. (\ref{inter}) has two points of intersection with a horizontal line $1/C^2$ if $C > e^{-1/2}$.}
\label{Fig2}
\end{figure}

{\it Turning points in a closed universe.}
Let us consider a closed relativistic universe, for which $k=1$.
The turning points for this universe are determined by the condition $\dot{y}=0$, for which Eq. (\ref{Jordan2}) gives
\begin{equation}
(3x^2 - 2x^4)e^{x^2}=\frac{1}{C^2}.
\label{inter}
\end{equation}
Figure \ref{Fig2} shows a curve representing the function on the left-hand side of Eq. (\ref{inter}).
This function has a maximum at $x=1$ and its value is $e$.
Equation (\ref{inter}) has solutions whose number depends on the value of $C$.
We only consider the solutions with a physical condition $x\ge0$.
This equation has two turning points (points of intersection of the curve in Fig. \ref{Fig2} with a horizontal line $1/C^2$) if the line lies below the maximum of the curve, that is, if
\begin{equation}
C > e^{-1/2}.
\label{condi}
\end{equation}
The segment of the curve above this line corresponds to $\dot{y}^2>0$ in Eq. (\ref{Jordan2}).
Consequently, the universe would oscillate between the two values of $x$ at the points of intersection: $x_\textrm{max}$ (big bounce) and $x_\textrm{min}$ (big crunch).
If $C=e^{-1/2}$, Eq. (\ref{Jordan2}) has one turning point at $x=1$.
The universe would be stationary at a constant temperature.
If $C<e^{-1/2}$, Eq. (\ref{Jordan2}) has no turning points and the universe would not exist.

Table~\ref{Table1} shows the values of $x_\textrm{min}$ and $x_\textrm{max}$ for a closed universe for different values of $C$.
As $C$ tends to infinity, the value of $x_\textrm{min}$ tends to $1/(\sqrt{3}C)$ which results from Eq. (\ref{inter}) reducing for small values of $x$ to $3x^2=1/C^2$.
In the same limit, the value of $x_\textrm{max}$ tends to $\sqrt{3/2}$ which is an $x$-intercept of the curve in Fig. \ref{Fig2} and sets the maximum temperature in the universe.

Table~\ref{Table1} also shows the values of $y$ corresponding to 
$x_\textrm{min}$ and $x_\textrm{max}$, given by Eq. (\ref{xy}), and the value of $y_\textrm{min}$, given by Eq. (\ref{mini}), for different values of $C$.
The value of $y(x_\textrm{max})$ is evidently greater than the corresponding value of $y_\textrm{min}$.
As $C$ tends to infinity, $y(x_\textrm{min})$ tends to $\sqrt{3}C^2$, whereas $y(x_\textrm{max})$ tends to $\sqrt{2/3}\,e^{3/4}C$.
In the same limit, the ratio $y(x_\textrm{min})/y(x_\textrm{max})$ tends to $(3/\sqrt{2})e^{-3/4}C$, $y(x_\textrm{min})/y_\textrm{min}$ to $\sqrt{3/e}C$, and $y(x_\textrm{max})/y_\textrm{min}$ to $\sqrt{2/3}e^{1/4}$.

\begin{table}[ht]
\begin{center}
\begin{small}
\begin{tabular}{ |c||c|c||c|c||c|}
\hline
$C$ & \quad $x_\textrm{min}$ & \quad $y(x_\textrm{min})$ & $x_\textrm{max}$ & \quad $y(x_\textrm{max})$ & \quad $y_\textrm{min}$ \\
\hline
$e^{-1/2}$ & 1 & 1 & 1 & 1 & 1\\
1 & 0.555209 & 2.10126 & 1.18912 & 1.70538 & $e^{1/2}$\\
10 & 0.057703 & 173.590 & 1.22444 & 17.2831 & 10$e^{1/2}$\\
100 & 0.005773 & 17320.9 & 1.22474 & 172.852 & 100$e^{1/2}$\\
\hline
\end{tabular}
\caption{The minima and maxima of $x$, the corresponding values of $y$, and the minimum of $y$, for different values of $C$ in a closed universe.
The domain of $C$ is $[1/\sqrt{e},\infty)$.}
\label{Table1}
\end{small}
\end{center}
\end{table}

{\it Turning points in a flat universe.}
Let us consider a flat relativistic universe, for which $k=0$.
The turning points for this universe are determined by the condition $\dot{y}=0$, for which Eq. (\ref{Jordan2}) gives
\begin{equation}
(3x^2 - 2x^4)e^{x^2}=0.
\end{equation}
This equation has two physical solutions (points of intersection of the curve in Fig. \ref{Fig2} with the $x$-axis).
They are given by $x_\textrm{max}=\sqrt{3/2}$ and $x_\textrm{min}=0$.
The segment of the curve above the $x$-axis corresponds to $\dot{y}^2>0$ in Eq. (\ref{Jordan2}).
Consequently, the temperature of the universe lies between $x_\textrm{min}$ and $x_\textrm{max}$.
As $x$ tends to 0, $y$ tends to infinity and $\dot{y}$, which is positive, asymptotically tends to zero.
Therefore, a flat universe has only one turning point at $x_\textrm{max}$.
This value coincides with that of $x_\textrm{max}$ for a closed universe in the limit $C\to\infty$.
The corresponding value of $y$ is $y(x_\textrm{max})=\sqrt{2/3}\,e^{3/4}C$ which is evidently greater than $y_\textrm{min}$ given by (\ref{mini}).

Table~\ref{Table2} shows the values of $x_\textrm{min}$ and $x_\textrm{max}$ for a flat universe for different values of $C$.
The values of $x_\textrm{min}$ and $x_\textrm{max}$ are $x$-intercepts of the curve in Fig. \ref{Fig2}.
Table~\ref{Table2} also shows the values of $y$ corresponding to $x_\textrm{min}$ and $x_\textrm{max}$, given by Eq. (\ref{xy}), and the value of $y_\textrm{min}$, given by Eq. (\ref{mini}), for different values of $C$.

\begin{table}[ht]
\begin{center}
\begin{small}
\begin{tabular}{ |c||c|c||c|}
\hline
$C$ & $x_\textrm{max}$ & \quad $y(x_\textrm{max})$ & \quad $y_\textrm{min}$ \\
\hline
1 & $(3/2)^{1/2}$ & $(2/3)^{1/2}e^{3/4}$ & $e^{1/2}$\\
10 & $(3/2)^{1/2}$ & 10$(2/3)^{1/2}e^{3/4}$ & 10$e^{1/2}$\\
100 & $(3/2)^{1/2}$ & 100$(2/3)^{1/2}e^{3/4}$ & 100$e^{1/2}$\\
\hline
\end{tabular}
\caption{The maximum of $x$, the corresponding value of $y$, and the minimum of $y$, for different values of $C$ in a flat universe.
Approximately, $x_\textrm{max}=1.22474$ and $y(x_\textrm{max})=1.72852\,C$.
The domain of $C$ is $(0,\infty)$.}
\label{Table2}
\end{small}
\end{center}
\end{table}

{\it Turning points in an open universe.}
Let us now consider an open relativistic universe, for which $k=-1$.
The turning points for this universe are determined by the condition $\dot{y}=0$, for which Eq. (\ref{Jordan2}) gives
\begin{equation}
(3x^2 - 2x^4)e^{x^2}=-\frac{1}{C^2}.
\end{equation}
This equation has one physical solution (point of intersection of the curve in Fig. \ref{Fig2} with a horizontal line $-1/C^2$).
It is given by $x_\textrm{max}$.
Therefore, an open universe has only one turning point at $x_\textrm{max}$.
The segment of the curve above this line corresponds to $\dot{y}^2>0$ in Eq. (\ref{Jordan2}).
Evidently, $x_\textrm{min}=0$.
As $x$ tends to 0, $y$ tends to infinity and $\dot{y}$, which is positive, asymptotically tends to 1.

Table~\ref{Table3} shows the values of $x_\textrm{min}$ and $x_\textrm{max}$ for different values of $C$.
As $C$ tends to infinity, the value of $x_\textrm{max}$ tends to $\sqrt{3/2}$, as for a closed universe, which is an $x$-intercept of the curve in Fig. \ref{Fig2} and equal to $x_\textrm{max}$ for a flat universe.
Table~\ref{Table3} also shows the values of $y$ corresponding to 
$x_\textrm{min}$ and $x_\textrm{max}$, given by Eq. (\ref{xy}), and the value of $y_\textrm{min}$, given by Eq. (\ref{mini}), for different values of $C$.
The value of $y(x_\textrm{max})$ is evidently greater than the corresponding value of $y_\textrm{min}$.
As $C$ tends to infinity, $y(x_\textrm{max})$ tends to $\sqrt{2/3}\,e^{3/4}C$, as for a closed universe.

\begin{table}[ht]
\begin{center}
\begin{small}
\begin{tabular}{ |c||c|c||c|}
\hline
$C$ & $x_\textrm{max}$ & \quad $y(x_\textrm{max})$ & \quad $y_\textrm{min}$ \\
\hline
0.01 & 2.33420 & 0.06531 & $e^{1/2}$\\
0.1 & 1.64821 & 0.23599 & $e^{1/2}$\\
1 & 1.25165 & 1.74866 & $e^{1/2}$\\
10 & 1.22505 & 17.2874 & 10$e^{1/2}$\\
100 & 1.22475 & 172.853 & 100$e^{1/2}$\\
\hline
\end{tabular}
\caption{The maximum of $x$, the corresponding value of $y$, and the minimum of $y$, for different values of $C$ in an open universe.
The domain of $C$ is $(0,\infty)$.}
\label{Table3}
\end{small}
\end{center}
\end{table}

{\it Time dynamics near a bounce.}
Differentiation of Eq. (\ref{xy}) with respect to $\tau$ gives
\begin{equation}
\dot{y}=C\dot{x}e^{x^2/2}\Bigl(1-\frac{1}{x^2}\Bigr).
\label{diff}
\end{equation}
Substitution of this $\dot{y}$ into Eq. (\ref{Jordan1}) gives
\begin{equation}
\dot{x}^2\Bigl(1-\frac{1}{x^2}\Bigr)^2+\frac{k}{C^2}e^{-x^2}=3x^2-2x^4,
\end{equation}
which determines the function $x(\tau)$, and with Eq. (\ref{xy}), the function $y(\tau)$.
Equation (\ref{diff}) shows that $\dot{y}=0$ if $\dot{x}=0$ or $x=1$.
As $x$ tends to 1, we have
\begin{equation}
\lim_{x\to 1}\dot{x}^2\Bigl(1-\frac{1}{x^2}\Bigr)^2=1-\frac{k}{eC^2}.
\end{equation}
Accordingly, as $x\to 1$, $\dot{x}\to\infty$: the function $x(\tau)$ has a vertical inflection point.

Equation (\ref{Jordan1}) for $x=1$, at which $y=Ce^{1/2}$ is a minimum of the function $y(x)$, gives
\begin{equation}
\dot{y}=\pm(eC^2-k)^{1/2}.
\end{equation}
This equation has solutions for all allowed values of $C$.
If $k=1$ and the condition (\ref{condi}) is satisfied, or if $k=0$ or $k=-1$, then $\dot{y}\neq 0$ at $x=1$.
It can be shown, by differentiation of Eq. (\ref{diff}) with respect to $\tau$, that the function $y(\tau)$ has a local minimum at $x=1$ and a local maximum when $\dot{x}=0$ ($x=x_\textrm{max}$).

Accordingly, the following scenario occurs.
As the universe contracts, $y(\tau)$ decreases and $x(\tau)$ increases.
When $x$ reaches 1, $y(\tau)$ reaches a local minimum (a bounce) and $x(\tau)$ exhibits vertical inflection.
Then, $x(\tau)$ continues to increase, but $y(\tau)$ also increases.
When $x$ reaches $x_\textrm{max}$, $y(\tau)$ reaches a local maximum.
Then, $x(\tau)$ begins to decrease, but $y(\tau)$ also decreases.
When $x$ reaches 1, $y(\tau)$ reaches a local minimum (another bounce) and $x(\tau)$ exhibits vertical inflection.
Then, $x(\tau)$ continues to decrease and $y(\tau)$ increases.
Consequently, at a bounce, the universe has a single bounce of the temperature and a double bounce of the scale factor with a little crunch between the two bounces.
If the value of $C$ remains constant, then the double bounce of the scale factor is symmetric.
Otherwise, it is asymmetric.

{\it Conclusions.}
We analyzed the dynamics of a homogeneous and isotropic universe in the Einstein--Cartan theory of gravity.
The coupling between the spin of fermions and torsion prevents a cosmological singularity, replacing the big bang with a nonsingular big bounce, at which the universe transitions from contraction to a short period of expansion, followed by a short period of contraction, followed by expansion (double bounce of the scale factor).
We eliminated the problem of a cusp that was reported earlier in \cite{Bounce}.
We showed that a closed universe may exist only when some function of the scale factor and temperature is higher than a particular threshold. 
Equation (\ref{xy}) and inequality (\ref{condi}) give
\begin{equation}
xy\,e^{-x^2/2}>e^{-1/2}.
\end{equation}
On the other hand, a flat universe and an open universe can exist for all positive values of the integration constant $C$.

The formation of a closed universe corresponds to the moment when the quantity $C$ begins to satisfy the inequality (\ref{condi}) in a region of space within a trapped null surface.
During inflation, this quantity increases \cite{ApJ}, possibly because of quantum particle-pair production in strong gravitational fields.
For large values of $C$, $x$ can become small when $y$ becomes large, as shown in Table~\ref{Table1}.
In this range, Eq. (\ref{xy}) reduces to the torsionless limit:
\begin{equation}
y(x)=\frac{C}{x},
\end{equation}
and the universe transitions from the torsion-dominated era to the radiation-dominated era.
The quantity $C$ begins to represent the product of the scale factor and temperature.
Eventually, the universe reaches the matter-dominated era and then dark-energy-dominated era.
Inflation must increase the value of $C$ to a threshold sufficient for the universe to reach the size at which dark energy can start the observed current acceleration \cite{ApJ}.
Otherwise, the universe would undergo several temperature bounces, accompanied by double scale factor bounces, before reaching this threshold \cite{SD}.

{\it Acknowledgment.}
This work was funded by the University Research Scholar program at the University of New Haven.

\end{document}